\documentstyle[pre,aps,multicol,epsfig]{revtex}

\begin{document} 



\title{Solitary Wave Interactions In Dispersive Equations Using
  Manton's Approach}
\author{P.G.\ Kevrekidis$^{1,3}$, Avinash Khare$^{2,3}$, and A. Saxena$^3$ } 
\address{
$^{1}$Department of Mathematics and Statistics, University of
Massachusetts, Amherst MA 01003-4515, USA \\
$^2$Institute of Physics,
Bhubaneswar, Orissa 751005, India \\
$^3$Theoretical Division and Center for Nonlinear Studies, Los
Alamos National Laboratory, Los Alamos, New Mexico 87545, USA}
\maketitle

\begin{abstract} 
We generalize the
approach first proposed by Manton [Nuc. Phys. B {\bf 150}, 397 (1979)]
to compute solitary wave interactions in translationally invariant,
dispersive equations that support such localized solutions. The
approach is illustrated using as examples solitons in the Korteweg-de
Vries equation, standing waves in the nonlinear Schr{\"o}dinger
equation and kinks as well as breathers of the sine-Gordon equation.
\end{abstract}

\vspace{3mm}

\begin{multicols}{2}

{\it Introduction}. 
Dispersive wave nonlinear partial differential equations (PDEs) 
describe a variety of  physical systems 
in atomic, optical, molecular, solid state and wave physics as well as
in fluid dynamics, biophysics,  
plasma physics, high energy physics 
and astrophysics among others \cite{general1,general2}.
Often in these settings, it is of particular interest to examine
the dynamics and interactions 
of spatially localized, and possibly travelling in time, 
solutions which can represent bits of information, moving Bose-Einstein
condensates, elementary particles or water waves 
\cite{soliton,infeld}. 
Particularly the interactions between the solitary waves are 
especially important, since, e.g., in optical communications 
avoiding such interactions  may reduce the bit error rate 
\cite{kivshar}; in 
Bose-Einstein condensates such interactions change significantly the
form of the wavefunction \cite{ketterle}, while in high energy  
physics models, the interactions are used to monitor the 
collisions of elementary particles \cite{anninos}.

There is a vast amount of literature regarding the 
interactions and collisions
of the above mentioned solitary waves. Many reviews of continuum
\cite{mal1,mal2} and discrete \cite{flach} systems contain some
of this enormous volume of work. The techniques that have been
used to identify the nature of such interactions are also rather
diverse ranging from perturbation theoretic ones as, e.g.,
in the work of Ref. \cite{karp}, to variational ones \cite{mal2,malmulti,ric},
to more rigorous ones, using the Fredholm alternative \cite{spiegel}
or Lin's method as in Ref. \cite{sand}. Typically, these
interactions asymptotically follow the tails of the
waves, which in most cases are exponential. This, in turn,
results in writing down Toda lattice type equations at the
``mesoscopic'' level for lattices of coherent nonlinear waves, see, e.g.,
Refs. \cite{ric,arnold} and references therein.

While the solitary wave interactions have been extensively studied 
in the past, in the present communication we aim at presenting a different
viewpoint on this topic. Our aim, in particular, is to explicitly focus on 
calculating the tail-tail interactions between the waves, using the
method proposed 
by Manton in Ref. \cite{manton}, based on the earlier work of Refs. 
\cite{goldberg,perring,rajaraman} and systematize it as a general method 
that can be straightforwardly applied to any nonlinear dispersive wave
equation that has a number of characteristics (which will be 
analyzed/explained). We believe that this method provides a
very simple, yet elegant and useful tool that can be generically used
in this large class of models and hence would be of value to 
researchers in a variety of disciplines.

The structure of our presentation is as follows: initially we
repeat Manton's formulation for the kink-antikink 
interaction in the sine-Gordon equation, highlighting some 
key and subtle points.
We then proceed to an analogous calculation for the case 
of solitons in the Korteweg-de Vries (KdV) equation. We then move
to the realm of breathers starting with the standing waves 
(with trivial periodicity) in the case of the
nonlinear-Schr{\"o}dinger (NLS) equation. Thereafter, we study 
the interaction of genuinely breathing structures such as the
breathers of the sine-Gordon equation. All of our results are  
corroborated with numerical simulations. Finally, we conclude
our presentation with a summary of main findings and some intriguing
questions for future study.

{\it Manton's Formulation for sine-Gordon Kinks}.
At the heart of Manton's calculation of the interaction energy
is the use of the definition of the linear momentum of the 
wave equation at hand. Computing the time derivative of the
momentum $P$ for an interval containing one solitary wave,
and deducing the contribution to it from the second wave,
we can infer the force exerted on the soliton from its neighbor.
In particular, for Lorentz-invariant equations (e.g. Klein-Gordon equations)
of the form:
\begin{eqnarray}
u_{tt}=u_{xx}-V'(u),
\label{sG}
\end{eqnarray}
the linear momentum can be written as:
\begin{eqnarray}
P=-\int u_t u_x dx.
\label{meq1}
\end{eqnarray}
The total integral along the line is a conserved
quantity (when the model is translationally invariant,
an assumption necessary for this approach to work).
Consider a soliton centered around $\xi=0$ ($u^{(1)}$),
and an antisoliton centered at $\xi=\Delta \xi$ ($u^{(2)}$), 
then using an interval
$(a,b)$ such that $a\ll0$ and $0\ll b\ll\Delta \xi$, we find that
in that interval (as shown in Ref. \cite{manton}):
\begin{eqnarray}
\frac{dP}{dt}=\left[-u_{x}^{(1)} u_x^{(2)} + u_{xx}^{(1)} u^{(2)} \right]_a^b.
\label{meq2}
\end{eqnarray}
If we then take into account the asymptotic form of the kink-like
waves, which for $V(u)=1-\cos(u)$ are of the form
$u =4 \arctan(e^{\pm x})$, we find that $u^{(1)} \approx 4 \exp(-x)$
and $u^{(2)} \approx 4 \exp(-(x-\Delta \xi))$ in the region between
the solitons (and hence at $x=b$). At $x=a$ the contribution
will be null as $a \rightarrow -\infty$. Manton's end result
is thus:   
\begin{eqnarray}
\frac{dP}{dt}=32 \exp(-\Delta \xi),
\label{meq3}
\end{eqnarray}
(which is {\it independent of b}) and hence the potential of
interaction is $V(\Delta \xi)=-32 \exp(-\Delta \xi)$.

There are some subtleties here that we would like to highlight: \\
$\bullet$ Notice that $dP/dt=\partial V/\partial \Delta \xi$ has been
used in the derivation of the potential, as opposed to the usual
$(-)$ sign on the right hand side. This is because, the way the
momentum is defined, $dP/dt>0$ which means that the momentum 
at $x=b$ is increasing, which in turn means that the solitary wave
is approaching $x=b$, because of the interaction and hence 
$\Delta \xi$ is decreasing (i.e., when $dP/dt>0$, the resulting
force in the dynamics of $\Delta \xi$ is negative). This sign
change accounts for the above formula used in the derivation of
the potential. This subtle point should always be taken into 
consideration when using Manton's method. \\
$\bullet$ Secondly, if we are to examine the dynamics of $\Delta \xi$,
then the corresponding Newton equation should account not only 
for the potential, but also for the ``mass'' of the solitary 
wave. In the case at hand (kinks of the sine-Gordon  equation), the
mass $M=\int u_x^2 dx=\int \left[ \frac{1}{2} u_x^2 + (1-\cos(u))
\right]$ is $8$.  \\
$\bullet$ Finally, for the dynamical equation of $\Delta \xi$, we 
should take into consideration that we have only computed the
force on the ``left soliton'' from the right one. However, 
there is an equal and opposite force on the right solitary
wave and hence we should use a factor of $2$ when writing
the equation for the acceleration of $\Delta \xi$. \\
Incorporating all these points in the equation for the separation
between the solitary waves we obtain that, Manton's formulation
yields in general:
\begin{eqnarray}
\ddot{\Delta \xi}= - \frac{2}{M} \frac{dP}{dt},
\label{meq4}
\end{eqnarray}
where the overdot denotes temporal derivative.
In the particular case of sine-Gordon kink-antikink interaction,
Eq. (\ref{meq4})
yields 
\begin{eqnarray}
\ddot{\Delta \xi}=-8 \exp\left(-\Delta \xi \right)\,.
\label{meq5}
\end{eqnarray}
A numerical example illustrating/validating the usefulness of this formula
is given in Fig. 1. 
It is worth remarking here that for all Lorentz-invariant theories, while
the kink-antikink interaction is attractive, the kink-kink interaction is
always repulsive.

\begin{figure}[htb]
\begin{center}
\includegraphics[scale = .375]{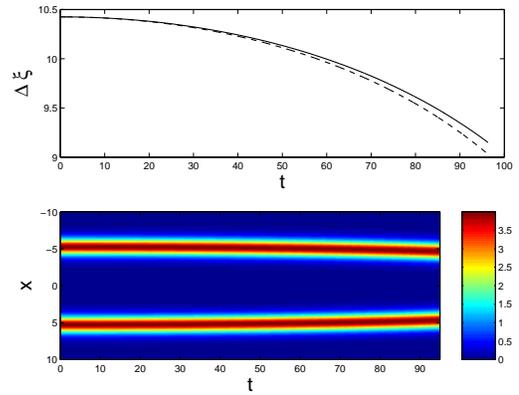}
\caption{Kink-antikink attractive interaction in the sine-Gordon 
equation. The kinks are initialized at a distance of $\approx 10.4$
and are allowed to interact. The top panel shows their separation
(numerically measured using a local center of mass approach weighted
on the energy density) as a function of time. The solid line shows
the numerical result, while the dashed one indicates the result of
integrating Eq. (\ref{meq5}).
The bottom panel shows
the space-time (x-t) contour plot of the energy density, indicating
the tendency of the waves to approach each other.}
\end{center}
\label{mfig1}
\end{figure}

{\it Soliton Interaction in the KdV Equation}.
Another interesting example that illustrates some additional
subtleties
of the Manton approach concerns the solitary wave interactions in
the KdV equation (most often associated with water waves 
\cite{soliton,infeld}). Following the same path as above, 
the definition of momentum in this case is given by $P=\int u^2 dx$,
hence we compute $dP/dt$. Given the form of KdV (in travelling wave frame
with velocity C):
\begin{eqnarray}
u_t=-u_{xxx} - 6 u u_x + C u_x\,,
\label{meq7}
\end{eqnarray}
we find that
\begin{eqnarray}
\frac{dP}{dt}=2 \int_a^b u u_t=\left[4 u^3-2 u u_{xx} +
  u_x^2 + C u^2 \right]_a^b.
\label{meq8}
 \end{eqnarray}
For a two-soliton decomposition $u=u^{(1)}+u^{(2)}$ and neglecting 
(exponentially smaller) higher order terms, we obtain the 
expression:
\begin{eqnarray}
\frac{dP}{dt} \approx 2 u_x^{(1)} u_x^{(2)} - 2 (u^{(1)} u_{xx}^{(2)} 
+ u^{(2)} u_{xx}^{(1)})
+ 2 C u^{(1)} u^{(2)}\,.
\label{meq9}
\end{eqnarray}
Finally, given the form of the soliton $u= (C/2) {\rm sech}^2(\sqrt{C}
x/2)$, the expression of Eq. (\ref{meq9}) is evaluated as:
\begin{eqnarray}
\frac{dP}{dt} = -16 C^3 \exp \left(-\sqrt{C} \Delta \xi \right). 
\label{meq10}
\end{eqnarray}
Using the expression of Eq. (\ref{meq4}), we would be immediately
inclined to write (given that the mass of the soliton is $\int u dx=2
\sqrt{C}$)
\begin{eqnarray}
\ddot{\Delta \xi}=16 C^{5/2} \exp\left(-\sqrt{C} \Delta \xi\right). 
\label{meq11}
\end{eqnarray}
However, we argue that when following Manton's formalism, one
should respect Ehrenfest's theorem. In particular, in the case
of KdV, it is true that:
\begin{eqnarray}
\frac{d}{dt} \int x u dx = 3 P \Rightarrow \frac{d^2}{dt^2} \int x u
dx = 3  \frac{d P}{dt}.
\label{meq12}
\end{eqnarray}
The left hand side indeed evaluates to $M \ddot{\Delta \xi}$
here (as well as for sine-Gordon), however the right hand side should be 
multiplied by a factor of 3. Hence, the resulting dynamical 
equation for the evolution of the soliton displacement should read:
\begin{eqnarray}
\ddot{\Delta \xi}=48 C^{5/2} \exp\left(-\sqrt{C} \Delta \xi\right).
\label{meq13}
\end{eqnarray}
A numerical example of the dynamics of KdV solitons is given in 
Fig. 2.

\begin{figure}[htb]
\begin{center}
\includegraphics[scale = .375]{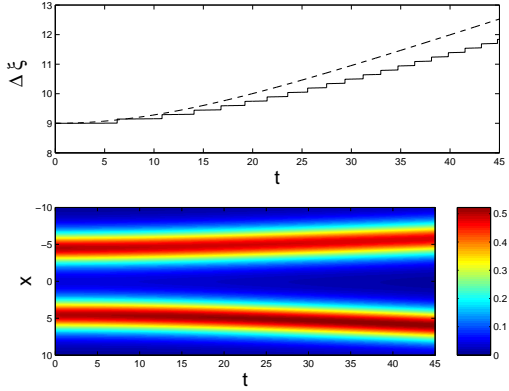}
\caption{The (repulsive) two-soliton interaction in the KdV equation.
The two panels are similar to those of Fig. 1. 
The 
small jumps in the top panel are due to the discretization used to emulate
the continuum equation. The two solitons are initialized at a distance
of $\Delta \xi=9$. The bottom panel shows the space-time (x-t) contour
plot of the field $u$.}
\end{center}
\label{mfig2}
\end{figure}

{\it Standing Wave Interaction in the NLS Equation}.
In the case of the NLS equation (a model relevant predominantly
to nonlinear optics and plasma physics 
\cite{kivshar}, but also more recently to  atomic physics as well
\cite{general2}), the role of the mass is played by the squared $L^2$ norm,
hence mass and momentum are defined as follows:
\begin{eqnarray}
M=\int |u|^2 dx; \hspace{5mm} P=\frac{i}{2} \int (u
u_x^{\star}-u^{\star} u_x) dx,
\label{meq14}
\end{eqnarray}
for the dynamical equation:
\begin{eqnarray}
i u_t=-\frac{1}{2} u_{xx} - |u|^2 u. 
\label{meq14a}
\end{eqnarray}
We can then compute similarly as above (up to higher order
exponentially smaller terms):
\begin{eqnarray}
\frac{dP}{dt}=\frac{1}{4} \left[u u_{xx}^{\star}+u_{xx} u^{\star} - 2
|u_x|^2 \right]_a^b.
\label{meq15}
\end{eqnarray}
We again use the standard 2-soliton decomposition, $u=u^{(1)}+u^{(2)}$ where
$u^{(1)}=\eta {\rm sech}(\eta x) \exp(i \eta^2 t/2)$, 
$u^{(2)}=\eta {\rm sech}(\eta (x-\Delta \xi)) \exp(i \eta^2 t/2) \exp(i \phi)$ 
are the standing waves, and the relative phase $\phi$ between them
has been incorporated in $u_2$. Then, one obtains:
\begin{eqnarray}
\frac{dP}{dt}=8 \eta^4 \exp \left(-\eta \Delta \xi \right),
\label{meq16}
\end{eqnarray}
which results in the dynamical equation for the separation
(using Eq. (\ref{meq4})) of the form:
\begin{eqnarray}
\ddot{\Delta \xi}=-8 \eta^3 \exp \left(-\eta \Delta \xi \right) \cos \phi\,.
\label{meq17}
\end{eqnarray}
This equation is identical to the one obtained in \cite{afa} 
through variational and perturbation techniques 
[cf. Eq. (5) of \cite{afa}].

We have also examined numerically the validity of 
the results. A typical computation is summarized in Fig. 3.
Notice that in the case of solitons in phase, the interaction
is attractive while it is repulsive when the solitons are ($\pi$) 
out of phase. This is also observed in the numerical experiments.

\begin{figure}[htb]
\begin{center}
\includegraphics[scale = .375]{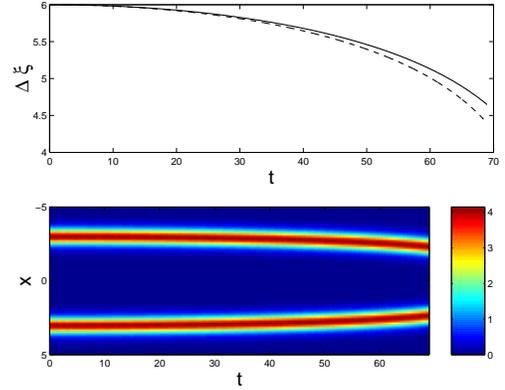}
\caption{Same as in Fig. 1,
but for the (in phase) standing waves
of the NLS equation. The latter were initialized at a distance of 6.
The bottom panel shows the space-time contour plots of the square
modulus of the wavefunction $u$.}
\end{center}
\label{mfig3}
\end{figure}

{\it Breather Interaction in the sine-Gordon Equation}. In the above  
examples we established cases where the interaction
between the solitary waves was previously known and used them
to analyze some of the key technical points 
of the Manton approach. We now turn to an example that, to the best
of our knowledge, has not been treated analytically before and which
concerns, in particular, the breather-breather interaction in the
sine-Gordon equation. Such solutions of Eq. (\ref{sG}) are given by 
the form:
\begin{eqnarray}
u=4 \arctan\left[\frac{\sqrt{1-\omega^2}}{\omega} \sin(\omega t) 
{\rm sech}\left(\sqrt{1-\omega^2} x\right) \right],  
\label{meq18}
\end{eqnarray}
where $\omega$ is the breather frequency.  A fundamental difference 
in this case is that the solutions are genuinely time dependent. 
However, the Manton formulation can still be carried through with 
the corresponding momentum derivative evaluated as:
\begin{eqnarray}
\frac{dP}{dt}=\left[-u_{t}^{(1)} u_t^{(2)} - u_{x}^{(1)} u_x^{(2)} 
+ (u_{xx}^{(1)} 
-u_{tt}^{(1)}) u_2 \right]_a^b.
\label{meq19}
\end{eqnarray}
As a result of the decomposition $u=u^{(1)}+u^{(2)}$ and the
asymptotics $u^{(1)} \approx 8\sqrt{1-\omega^2} 
\sin(\omega t) \exp \left(-\sqrt{1-\omega^2}
x\right)/\omega$ and 
$u^{(2)} \approx 8\sqrt{1-\omega^2} \sin(\omega t)  
\exp\left(-\sqrt{1-\omega^2}
(x-\Delta \xi)\right)/\omega$, we obtain:
\begin{eqnarray}
\frac{dP}{dt}=\frac{64 (1-\omega^2)^{2}}{\omega^2} 
\left(1-\frac{1}{1-\omega^2} \cos(2 \omega t)\right) \nonumber \\ 
\times\exp(-\sqrt{1-\omega^2} \Delta \xi).  
\label{meq20}
\end{eqnarray}
From this calculation, we can infer that while the potential
of interaction between two sine-Gordon breathers is non-autonomous, it 
has a well defined attractive (when the breathers are in phase)
average of $V=-(64/\omega) (1-\omega^2)^{3/2} 
\exp(-\sqrt{1-\omega^2} \Delta \xi)$. Furthermore, using Eq. (\ref{meq4})
and the breather mass $M=16 \sqrt{1-\omega^2}$, we can infer the
dynamical equation governing the inter-breather separation as:
\begin{eqnarray}
\ddot{\Delta \xi}=-\frac{8 (1-\omega^2)^{3/2}}{\omega^2} 
\left(1-\frac{1}{1-\omega^2} \cos(2 \omega t)\right) \nonumber \\ 
\times\exp(-\sqrt{1-\omega^2} \Delta \xi). 
\label{meq21}
\end{eqnarray}

A numerical experiment illustrating the comparison of
Eq. (\ref{meq21}) with the PDE result
is shown in Fig. 4.

\begin{figure}[htb]
\begin{center}
\includegraphics[scale = .375]{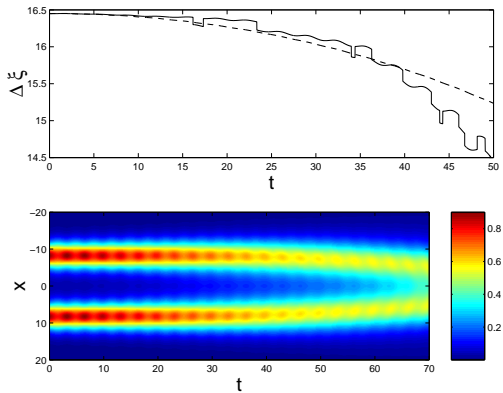}
\caption{Same as in Fig. 1,
but for the (attractive)
  interaction of in-phase breathers of the sine-Gordon equation.
The breathers are initialized at a distance of $\approx 16.45$.
The dashed line is the theoretical prediction of Eq. (\ref{meq21}).
The small jumps in the numerical result are an artifact of the
discretization used to approximate the continuum equation. However,
it can be clearly seen that the theoretical prediction follows
well the full numerical evolution at least for times $t<35$, 
where the breathers are sufficiently well separated that the
interaction has not changed their profile significantly (and for
which they maintain their individual character).}
\end{center}
\label{mfig4}
\end{figure}

{\it Conclusions}. In this short communication, we revisited the
topic of solitary wave interactions for exponentially localized
solutions (even though that is not absolutely necessary; however,
it is typically the case) of translationally invariant, nonlinear
dispersive wave equations. We showed  how to systematically 
exploit the momentum integral to find the force exerted on one
of the waves by the other and how to establish the dynamical equation of motion
of the inter-soliton distance, using the mass integral and Ehrenfest's
theorem [notice that due to the structure of the energy-momentum tensor
for Lorentz invariant equations, the analog of
such a theorem in the latter case is given by $d/dt \int x {\cal E}
dx=P$, where ${\cal E}$ is the energy density]. 
We demonstrated the use of the approach in a number of
well-established situations, including the kink interaction in the
sine-Gordon equation, the soliton interaction in the Korteweg-de Vries
equation and the standing wave dynamics in the nonlinear
Schr{\"o}dinger
equation. The method was also applied to obtain an analytical expression
for the breather-breather interaction in the sine-Gordon model.

Naturally, the approach has a number of limitations: for
instance it cannot be directly applied in the presence of spatially
dependent potentials. Similarly, it cannot be straightforwardly 
implemented in cases where Ehrenfest's theorem (or Lorentz invariance)
cannot be 
established. A prominent such example is, for instance, the modified
Korteweg-de Vries equation wherein $d/dt \int x u dx =2 \int u^3 dx$,
whereas the momentum is given by $\int u^2 dx$. Certainly, it is also
of interest to generalize the approach to multiple dimensions in
the spirit of Ref. \cite{malmulti}. Finally, an alternative method
to compute the interaction energy of solitary waves is through 
direct energy arguments: in particular, if the corresponding  
soliton lattice solution exists, one can compute the energy of 
such a solution, inside a period.
The leading term will then be
the energy of a single soliton, while the corrections will 
correspond to the soliton interaction energy. It would
be interesting to compare/validate the results of such a method
with those of Manton's formalism. These topics will be deferred to 
future publications.

{\it Acknowledgements}. 
PGK acknowledges the support of NSF-DMS-0204585, NSF-CAREER
and the Eppley Foundation for Research. Research at Los Alamos
was supported by the U.S. Department of Energy.

\end{multicols}


\begin{references}

\vspace{-15mm}

\bibitem{general1} M. Remoissenet, {\it Waves called Solitons} 
(Springer-Verlag, Berlin, 1999). 

\bibitem{general2} L. P. Pitaevskii and S. Stringari, 
{\it Bose Einstein Condensation} (Clarendon Press, Oxford, 2003);
R.K. Dodd, J.C. Eilbeck, J.D. Gibbon, and H.C. Morris, 
{\it Solitons and Nonlinear Wave Equations} 
(Academic Press, London, 1982). 

\bibitem{soliton} P.G. Drazin, {\it Solitons} (Cambridge University
Press, Cambridge, 1983).


\bibitem{infeld} E. Infeld and G. Rowlands, 
{\it Nonlinear Waves, Solitons and Chaos} (Cambridge University Press, 
Cambridge, 2000).

\bibitem{kivshar} Y. Kivshar and G.P. Agrawal,
{\it Optical Solitons: From Fibers to Photonic Crystals}
(Academic Press, San Diego, 2003).

\bibitem{ketterle} J.M. Vogels, J.K. Chin, and W. Ketterle, 
Phys. Rev. Lett. {\bf 90}, 030403 (2003). 

\bibitem{anninos} P. Anninos, S. Oliveira, and R. A. Matzner, 
Phys. Rev. D {\bf 44}, 1147 (1991).

\bibitem{mal1} Yu.S. Kivshar and B.A. Malomed,
Rev. Mod. Phys. {\bf 61}, 763 (1989).  

\bibitem{mal2} B.A. Malomed, Progress in Optics {\bf 43}, 69 (2002).

\bibitem{flach} P.G. Kevrekidis, K.{\O}. Rasmussen, and A.R. Bishop,
Int. J. Mod. Phys. B {\bf 15}, 2833 (2001); S. Flach and C.R. Willis,
Phys. Rep. {\bf 295}, 181 (1998).

\bibitem{karp} V.I. Karpman and V.V. Solov'ev, Phys. D {\bf 3}, 142 
(1981); V.I. Karpman and V.V. Solov'ev, Phys. D {\bf 3}, 487 (1981).

\bibitem{malmulti} B.A. Malomed, Phys. Rev. E {\bf 58}, 7928 (1998).

\bibitem{ric}  R. Carretero-Gonz{\'a}lez and K. Promislow, 
Phys. Rev. A {\bf 66},  033610 (2002).

\bibitem{spiegel} C. Elphick, E. Meron, and E.A.  Spiegel, 
SIAM J. Appl. Math. {\bf 50}, 490 (1990).

\bibitem{sand} B. Sandstede,
Trans. Amer. Math. Soc. {\bf 350}, 429 (1998).

\bibitem{arnold} J. M. Arnold, Phys. Rev. E {\bf 60}, 979 (1999).

\bibitem{manton} N.S. Manton,
Nucl. Phys. B {\bf 150}, 397 (1979).

\bibitem{goldberg} J.N. Goldberg, P.S. Jang, S.Y. Park, and K.C. Wali,
\newblock Phys. Rev. D {\bf 18}, 542 (1978).

\bibitem{perring} J.K. Perring and T.H.R. Skyrme, \newblock Nucl. Phys.
{\bf 31}, 550 (1962).

\bibitem{rajaraman} R. Rajaraman, \newblock Phys. Rev. D {\bf 15}, 2866
(1977).


\bibitem{afa} V.V. Afanasjev, B.A. Malomed, and P.L. Chu,
Phys. Rev. E {\bf 56}, 6020 (1997).





\end{references}
\end{document}